\documentclass[twocolumn,showpacs,pre,letterpaper,final]{revtex4}
\usepackage{amscd}
\usepackage{amssymb}
\usepackage{amsfonts}
\usepackage{amsmath}
\usepackage{graphicx}
\usepackage{dcolumn}
\usepackage{bm}
\usepackage{textcomp}
\usepackage{multirow}
\usepackage[dvipdfm]{hyperref}

\hypersetup{colorlinks=true, linkcolor=blue, anchorcolor=blue,
citecolor=blue, filecolor=blue, pagecolor=blue, urlcolor=blue}
\begin{document}

\preprint{}
\title{Pairwise interaction pattern in the weighted communication network}
\author{Xiao-Ke Xu$^{1,2}$}
\email{xiaokeeie@gmail.com}
\author{Jian-Bo Wang$^{3}$}
\author{Ye Wu$^{4}$}
\author{Michael Small$^{5,1}$}
\email{small@ieee.org}

\affiliation{$^1$Department of Electronic and Information
Engineering, Hong Kong Polytechnic University, Hong Kong \\
$^2$School of Communication and Electronic Engineering, Qingdao Technological University, Qingdao 266520, China\\
$^3$Adaptive Networks and Control Lab, Department of Electronic
Engineering, Fudan University, Shanghai 200433, China\\
$^4$ School of Science, Beijing University of Posts and
Telecommunications, Beijing 100876, China \\$^5$School of
Mathematics and Statistics, University of Western Australia,
Crawley, WA 6009, Australia}

\date{\today}

\begin{abstract}
Although recent studies show that both topological structures and
human dynamics can strongly affect information spreading on social
networks, the complicated interplay of the two significant factors
has not yet been clearly described. In this work, we find a strong
pairwise interaction based on analyzing the weighted network
generated by the short message communication dataset within a
Chinese tele-communication provider. The pairwise interaction
bridges the network topological structure and human interaction
dynamics, which can promote local information spreading between
pairs of communication partners and in contrast can also suppress
global information (e.g., rumor) cascade and spreading. In addition,
the pairwise interaction is the basic pattern of group conversations
and it can greatly reduce the waiting time of communication events
between a pair of intimate friends. Our findings are also helpful
for communication operators to design novel tariff strategies and
optimize their communication services.
\end{abstract}


\maketitle

\section{Introduction}
Connections based on both information transmission and human
relationships coexist in different types of electronic
person-to-person communications, like in e-mail
networks\cite{Eckmann_2005, Vazquez_2007, Iribarren_2009}, instant
message services \cite{Leskovec_2008}, mobile phone calls
\cite{Onnela_2007, Karsai_2011, Miritello_2011}, and mobile phone
short messages \cite{Hong2009shortmessage, Wu_2010,
Zhao2011shortmessage}. Basically, there are two significant factors
affecting spreading processes on social contact networks:
topological structures \cite{Pastor-Satorras_2001, Small_2007} and
human dynamics \cite{Vazquez_2007, Iribarren_2009, Barabasi_2005}.
Previous studies have uncovered how different topological structures
(such as degree distribution \cite{Barabasi_1999,
Pastor-Satorras_2001}, assortativity \cite{Newman_2002,
Boguna_2003}, modularity \cite{Newman_2006, Zhang_2010}) affect
spreading dynamics. At the same time, it has been proved that the
non-Poissonian pattern of human dynamics can slow down information
propagation \cite{Vazquez_2007, Iribarren_2009, Miritello_2011,
Min_2011}.

Although the above studies give us valuable insights into how
information spreads on human communication networks, they do not
provide a unified framework to combine the effects of topological
structures and human dynamics on information diffusion. To uncover
the complicated interplay between network structures and human
dynamics, recently the concept of temporal networks has been
proposed \cite{Holme_2011} and related statistics have been utilized
to analyze human communication networks \cite{Pan_2011}. For
example, Miritello and collaborators show that group conversations
are significant to understand the dynamic coupling of individuals
and the mechanism of information spreading \cite{Miritello_2011}.
Furthermore, some basic statistics for local group conversations,
such as temporal motif \cite{Zhao_2010, Kovanen_2011} and weighted
reciprocity \cite{Kovanen2010reciprocity, Chawla2011reciprocity,
Cheng2011reciprocity}, have been utilized to measure how individuals
interact with their direct neighbors. However, up to now, what the
accurate interaction pattern among individuals in electronic
person-to-person communications is and how the interaction pattern
affects spreading behaviors has not been satisfactorily uncovered.

In recent years, the mobile short message service has emerged as one
of the most popular tools for personal communication in China
because of the relatively low telecommunication tariff
\cite{Hong2009shortmessage, Wu_2010, Zhao2011shortmessage}.
Basically, people can send and receive short messages anytime,
anywhere, and to and from any mobile phone in their daily life. In contrast,
e-mails and instant message services (e.g., MSN) are not as popular
as short messages in China for they need people to have a personal
computer or a mobile smart phone. Furthermore, handling a short
message is easier than writing an e-mail, and is more flexible than
dialing a phone call. The flexibility of short messages means that
you can send or respond a short message promptly, or put some
particular short messages to a waiting list as a lower priority task
\cite{Wu_2010}. These features thus provide a very attractive proxy
for studying the dynamic behaviors of single individuals and the
nontrivial interaction pattern of human communication activities.

In this study, we build a weighted communication network based on a
short message dataset of a telecommunication provider in China. By
analyzing the network topology and interaction strength, we find
that most users have only one major active communication partner.
Furthermore, the results of multiple statistics indicate that the
weighted communication network has specific structural features
compared with its randomized counterpart: densely mutual
(reciprocal) links, highly weighted reciprocal coefficients
\cite{Kovanen2010reciprocity}, and fragmented pairwise rich nodes.
All these topology properties can prove the existence of the
pairwise interaction pattern in the weighted network, which has not
been reported in previous studies.

The pairwise pattern shows that most information in the short
message service is local, especially tends to happen between a pair
of users. Therefore, the pairwise interaction is the basic pattern
of group conversations in the self-organizing artificial
communication system (e.g., mobile phone short message service).
This finding is significant for our understanding of local and
global information spreading in electronic person-to-person
communications. We also find the pairwise conversation can strongly
affect human dynamics in the short message service. Moreover, our
study provides an integrated framework for analyzing the collective
communication behaviors in the self-organizing communication system
based on weighted network theory. Therefore, communication operators
can use our findings to design novel service plans, provide new
tariff strategies, and optimize the communication services.

\section{Short message network}
\subsection{Degree and weight distributions}

The short message dataset investigated in this work was obtained
from a mobile phone operator in China. The original data includes
all the charging accountant bills over one month period for three
corporate users who use the same mobile phone operator service. In
this study, we only calculate the result for the company that has
the most number of users. In this dataset, the total number of the
short message communication records is $643,502$, and the number of
the users is $72,146$. Each record comprises a sender mobile phone
number, a recipient mobile phone number and a time stamp with a
precision of $1$ s. All the phone numbers were hashed and no other
information is available for identifying or locating users. More
detailed information about this dataset can be found in
\cite{Wu_2010}.

\begin{figure}[htbp]
\centering
\includegraphics[width=0.5\textwidth]{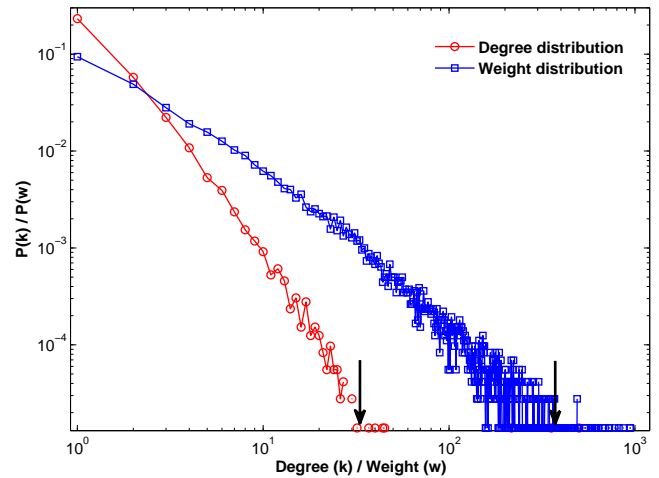}
\caption{(Color online) Degree and weight distributions for the
short message communication network. The dataset includes all the
charging accountant bills over one month period for the users who
use the same mobile phone operator service in a company. The total
number of the short message communication records is $643,502$, and
the number of the users is $72,146$.} \label{degree_distribution}
\end{figure}

To show the interaction pattern among the communication network
neighbors, we build a weighted complex network representing the
communication relationships among the members of the company. To
eliminate spurious relationships, we only retain the links with
bidirectional short message communication within the largest
connected component (as has been done for mobile phone call data in
\cite{Karsai_2011}), thus the out-degree always equals the
in-degree. The link weights are the number of short message events.
We list the number of friends (degree) distribution and the number
of short messages (weight) distribution in Fig.
\ref{degree_distribution}. The degree and weight distributions are
both like a power-law distribution \cite{Barabasi_1999}, which
suggests that a very large percentage of users have small values of
degree/weight, and few users possess very high degrees/weights.
However, the values of natural cutoff for the degree and weight are
very different: $k_{c}\approx 30$ and $w_{c}\approx 400$. The result
of $w_{c}\gg k_{c}$ indicates that the weight distribution is more
heterogeneous than the degree distribution.

\subsection{Uncorrelation of degree and weight}
\begin{figure}[htbp]
\centering
\includegraphics[width=0.5\textwidth]{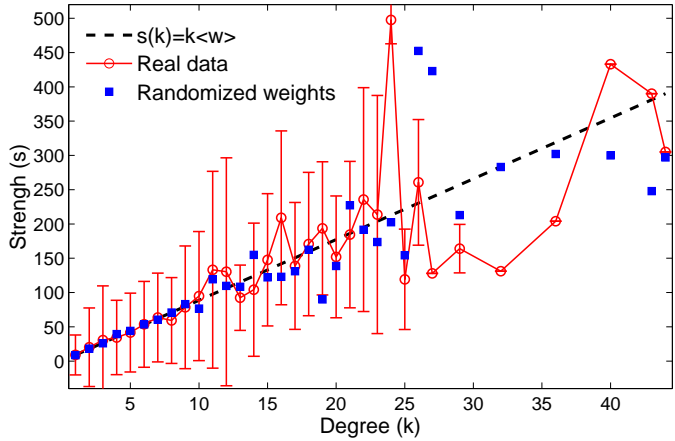}
\caption{(Color online) The out-strength $s(k)$ is an uncorrelated
function of the node degree $k$: $s(k)=\langle w\rangle k$
\cite{Barrat_2004}. The value of $s(k)$ is the averaging
out-strength $s_i$ for the nodes with the same degree $k$, and the
standard deviation is plotted. The result of the real data is very
similar to that obtained in a randomized weighted network. We
maintain the topology of the original network and shuffle the
weights globally to get the randomized network
\cite{Opsahl_2008richclub}.} \label{strength_degree}
\end{figure}

We extend the vertex degree $k_i$ to be a new definition of the
vertex out-strength as $s_i=\sum _{j\in N} w_{ij}$
\cite{Barrat_2004}. To shed more light on the relationship between
the vertices' strength and degree, we investigate the dependence of
the average strength $s(k)$ with the degree $k$ increasing. If there
is no correlation between the weight and the degree, we can obtain
$s(k)=\langle w\rangle \sum _{j}a_{ij}=\langle w\rangle k$, where
$\langle w\rangle$ is the average out-weight in the network
\cite{Barrat_2004}. We show the results of $s(k)$ for both the real
weighted network and its randomized version in Fig.
\ref{strength_degree}. To get the randomized network, we maintain
the topology of the original network and shuffle the weights
globally to remove  possible correlation of the weight and degree
\cite{Opsahl_2008richclub}. The two curves are very similar and both
closely fitting the relationship: $s(k)=\langle w\rangle k$. The
strength of a vertex is simply proportional to its degree, which
means that link weights are independent upon node degrees.

\begin{figure}[htbp]
\centering
\includegraphics[width=0.5\textwidth]{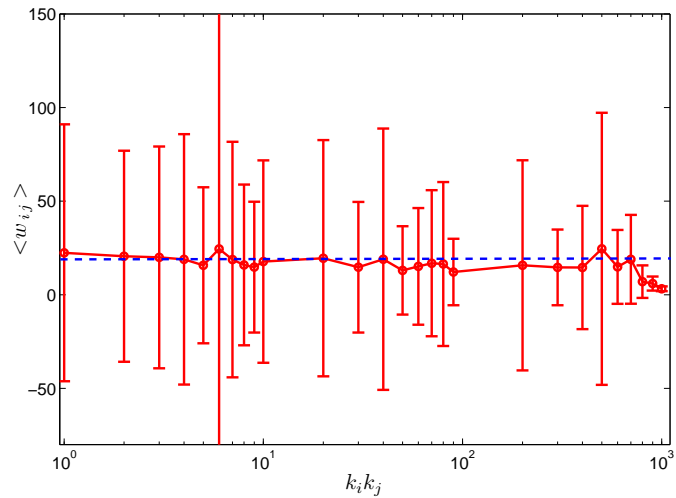}
\caption{(Color online) The weight $\langle w_{ij}\rangle$ is almost
constant for about three decades of $k_ik_j$, where $k_i$ and $k_j$
are the end-point node degrees. The value of $\langle w_{ij}\rangle$
is the averaging result for the links with the same $k_ik_j$, and
the standard deviation is plotted. The result confirms that there
are no correlations between the weight and two corresponding
end-point node degrees.} \label{weight_kikj}
\end{figure}

Another method of measuring the correlation between the degree and
weight is to calculate the dependence of the weight $w_{ij}$ on the
degrees of the end-point node degrees $k_i$ and $k_j$
\cite{Barrat_2004}. As we can see in Fig. \ref{weight_kikj}, the
average weight $\langle w_{ij}\rangle$ is almost constant for about
three decades of $k_ik_j$, confirming a general lack of correlations
between the weight and the end-point node degrees. All the above
results imply that the weight and degree have fundamentally
different coupling rules, so we need to consider not only the
network topology, but also interaction strengths among nodes in the
following sections.

\section{Pairwise interaction pattern}
\subsection{Heterogeneous communication pattern}
\begin{figure}[htbp]
\centering
\includegraphics[width=0.5\textwidth]{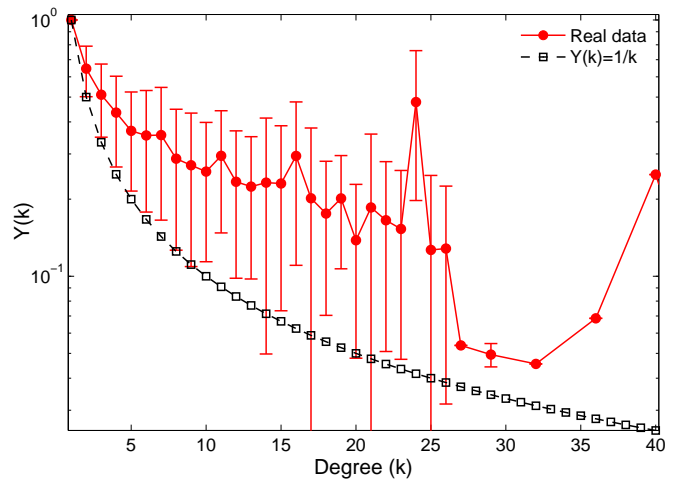}
\caption{(Color online) Measured $Y(k)$ as a function of $k$ is
shown and compared with its theoretical low-boundary $1/k$ and
upper-boundary $1.0$. The value of $Y(k)$ is the averaging $Y_i$ for
the nodes with the same degree $k$, and the standard deviation is
plotted. The value of $Y(k)$ is the disparity averaged over all
nodes with the degree $k$ \cite{Almaas_2004}. Because the curve of
$Y(k)$ is far larger than $1/k$, the intensity of the users'
communication is highly heterogeneous among their network
neighbors.} \label{Yk_degree}
\end{figure}

For a given node $i$ with degree $k_i$ and out-strength $s_i$, the
disparity in the weights can be evaluated by the quantity $Y_i=\sum
_{j\in N_i} \left [w_{ij}/s_i \right ]^2$, where $N_i$ is the set of
first neighbors of $i$ \cite{Derrida_1987, Barthelemy_2003}. By this
definition, $Y_i$ has an implicit dependence on the value of $k_i$.
If all edges have comparable weights, $Y(k)$ (that is, the disparity
averaged over all nodes with the degree $k$) will be scaled as
$1/k$. In other words, the value of the weight $w_{ij}$ is of the
same order $s_i/k_i$. In contrast, if only one or a few weights
dominate over all the others, $Y(k)$ is independent of $k$ and
$Y(k)\simeq1.0$ \cite{Almaas_2004}. The result of $Y(k)$ for the
short message communication network has been shown in Fig.
\ref{Yk_degree}. It is clear that the curve of $Y(k)$ is far larger
than $1/k$, which suggests that the users show a strong
heterogeneous communication pattern. This observation proves that
most users mainly have heavy communication with just one of their
friends \cite{Wu_2010}.

\subsection{Principal communication component}
\begin{figure}[htbp]
\centering
\includegraphics[width=0.5\textwidth]{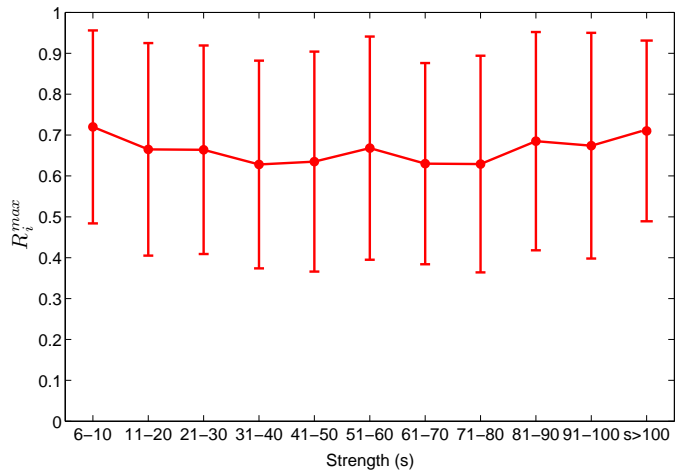}
\caption{(Color online) The values of $R_i^{max}$ for active users.
The active users are the persons who send more than five short
messages. We see that about $70\%$ of short messages from each user are
sent to one particular associate, which clearly reveals that most of
the users indeed have only one major active communication partner.}
\label{rsmax}
\end{figure}

In the previous study, it has been found that about $50\%$ of the
users send more than $90\%$ of their messages to just one friend
\cite{Wu_2010}. We define the statistic $R_{ij}=n_{ij}/N_i$ to
measure the intensity of communication between users $i$ and $j$.
The value of $R_{ij}$ is the ratio of the number of message $n_{ij}$
that $i$ sends to $j$ to the total messages $N_i$ sent by $i$. For
user $i$, it is very easy to find the maximal value $R_i^{max}$ from
his short message records. The value of $R_i^{max}$ close to $1.0$
indicates that the user mainly communicates with one particular
partner. For many individuals who only send one or two short
messages, the values of $R_i^{max}$ are obviously very high ($1.0$
or $0.5$). Therefore, we only list the values for active users who
send more than five short messages in Fig. \ref{rsmax}. The result
of $R_i^{max}$ is stable for all the active users (meaning about
$70\%$ messages are sent to the most closest person), which clearly
reveals that most of the users indeed have only one major active
communication partner.

\subsection{Reciprocity of communication}
Our data analysis also shows that mutual (reciprocal) links are
dense in the communication network. A traditional way of quantifying
the reciprocity is to compute the ratio of the number of links
pointing in both directions $L^\leftrightarrow$ to the total number
of links $L$: $r=\frac{L^\leftrightarrow}{L}$
\cite{Definition_reciprocity,Pattern_reciprocity}. The value of $r$
for a real network lies in the range of $[0,1]$. However, this
definition is not very useful for the weighted network built by the
short message communication events, for we have removed all
unidirectional edges and the reciprocity according to the above
definition would be $1.0$.

\begin{figure}[htbp]
\centering
\includegraphics[width=0.5\textwidth]{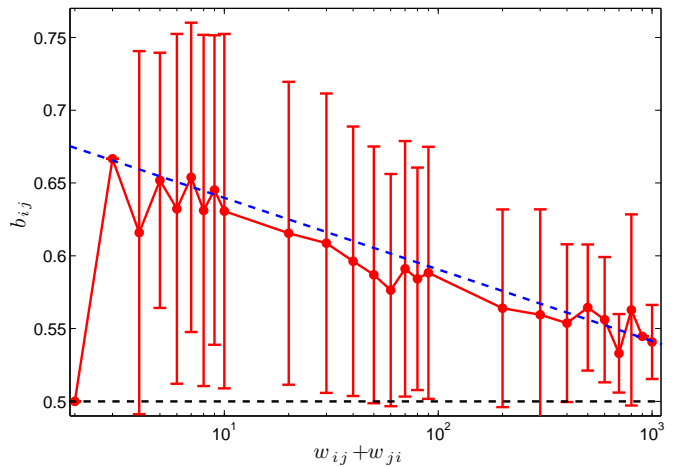}
\caption{(Color online) The values of weighted reciprocal
coefficients $b_{ij}$ as a function of edge weights for all the
users. The weighted reciprocal coefficient is defined as
$b_{ij}=max(\frac{w_{ij}}{w_{ij}+w_{ji}},
\frac{w_{ji}}{w_{ij}+w_{ji}})$, where $w_{ij}$ is the total number
of short messages that user $i$ sends to $j$
\cite{Kovanen2010reciprocity}. All the values of $b_{ij}$ are below
$0.7$ and decrease with $w_{ij}+w_{ji}$ increasing, which indicates
that the information flow has a symmetric (reciprocal) trend in the
short message commutation network. The blue dash line is just drawn
to guide our eyes.} \label{reciprocity}
\end{figure}

For a weighted network, we aim to determine not only whether the
communication between a pair of users is one directional or mutual
\cite{Cheng2011reciprocity}, but also the bias of the information
flow strength of the two different directions
\cite{Chawla2011reciprocity, Kovanen2010reciprocity}. Here we use
the definition in \cite{Kovanen2010reciprocity}:
$b_{ij}=max(\frac{w_{ij}}{w_{ij}+w_{ji}},
\frac{w_{ji}}{w_{ij}+w_{ji}})$. The weight $w_{ij}$ is the total
number of short messages that user $i$ sends to $j$ during the whole
month period. Note that $b_{ij}$=$b_{ji}$, and the distributions of
$b_{ij}$ for all edges in the network are in the range of
$[0.5,1.0]$. The weight of each edge allows us to study the
reciprocity of the edge instead of the full network. The value of
$0.5$ means that the edge is symmetric (reciprocal), and in contrast
the value of $1.0$ means that the information flow is completely
unidirectional.

The dependence of $b_{ij}$ on the value of $w_{ij}+w_{ji}$ is shown
in Fig. \ref{reciprocity}. This result is an obvious evidence that
there is a reciprocal trend for short message communications, for
all the values of $b_{ij}$ are below $0.7$. In addition, the more
reciprocity property for the higher $w_{ij}+w_{ji}$ also implies
that a more symmetric pairwise relationship might trigger more
communication events, and vice versa.

\subsection{Pairwise structure of rich nodes}
\begin{figure}[htbp]
\centering
\includegraphics[width=0.5\textwidth]{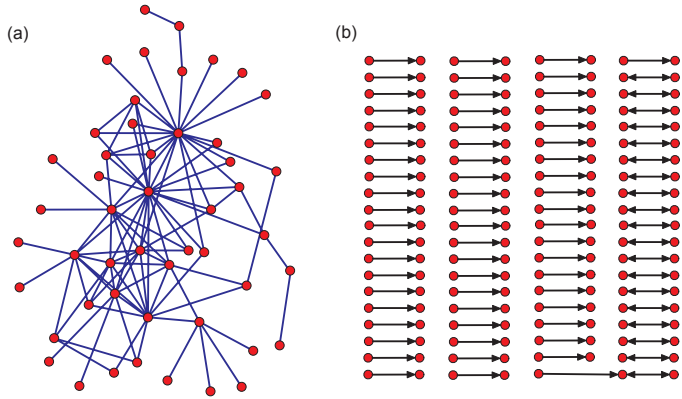}
\caption{(Color online) The topological structure of the edges with the
$100$ highest weight in the air transportation network and the short
message communication network. To simplify the representation of the
structure, here we do not show the weight of each edge. Compared
with the strong inter-connection of the air transportation network,
the short message communication network shows many isolated pairwise
connections. (a) The air transportation network is obtained by
considering the $500$ United States airports with the largest amount
of traffic from publicly available data \cite{Colizza_2007}. Nodes
represent airports and edges represent air travel connections among
them. The weights are given by the number of seats available on the
scheduled flights. (b) The short message communication network is
obtained by considering the short message sending/receiving
relationships among the members of a company in China
\cite{Wu_2010}. Nodes represent individuals and edges represent
short message communication among them. The weights are given by the
number of short messages.} \label{multipolarization}
\end{figure}

The rich-club phenomenon is another significant topology property
\cite{Richclub_origin, Colizza_richclub} and it can dominate other
statistics of complex networks \cite{Xu_2010richclub,
Xu_2011richclub}. In many real-life weighted networks, the rich-club
effects based on degree and strength are not trivially related
\cite{Opsahl_2008richclub, Serrano_2008}. The topology structure of
the $100$ highest weight edges in the short message communication
network is shown in Fig. \ref{multipolarization}(b). We also compare
this result with that of the U.S. air transportation network
\cite{Colizza_2007}. In the air transportation network, the $100$
highest weight edges connect each other and form an inter-connected
group [Fig. \ref{multipolarization}(a)]. Compared with the tendency
of the rich-club effect in the air transportation network, the short
message communication network shows many isolated pairwise
connections, so our results illustrate that the short message
communication network has no rich-club property.

Although we do not calculate the rich-club coefficient using
quantitative methods \cite{Richclub_origin, Colizza_richclub,
Opsahl_2008richclub, Serrano_2008}, the very fragmented local
subsets again prove the pairwise interaction pattern in the short
message communication network. Our simply qualitative method not
only measures the relationship among the edges with the highest
weights but also roughly represents the connection of the nodes with
the strongest strengths. Because all users send most of their short
messages to one major active communication partner [Fig.
\ref{rsmax}], the nodes with the strong strengths in the short
message communication network have a similar connection relationship
like Fig. \ref{multipolarization}(b).

\section{Impact of pairwise interaction pattern}
We have known the accurate interaction pattern among individuals,
and the next step is to explore how the interaction pattern affects
spreading behaviors in electronic person-to-person communications.

\subsection{Impact on information spreading}
To show how the pairwise interaction pattern affects spreading
behaviors, we use the Susceptible-Infected (SI) model to simulate
information spreading dynamics \cite{Onnela_2007, Karsai_2011}. For
the weighted short message network, each infected individual $i$ can
pass the information to his direct neighbor $j$ with the probability
$P_{ij}=xw_{ij}$ at each time step, where the parameter $x$ controls
the overall spreading rate \cite{Onnela_2007}. We also compare the
above result with that obtained in an unweighted version (all the
tie weights are considered equal) of the short message communication
network using the same average spreading rate. Recently, it has been
found that information transferring is significantly slower in the
weighted network than in its unweighted version \cite{Onnela_2007}.

\begin{figure}[htbp]
\centering
\includegraphics[width=0.5\textwidth]{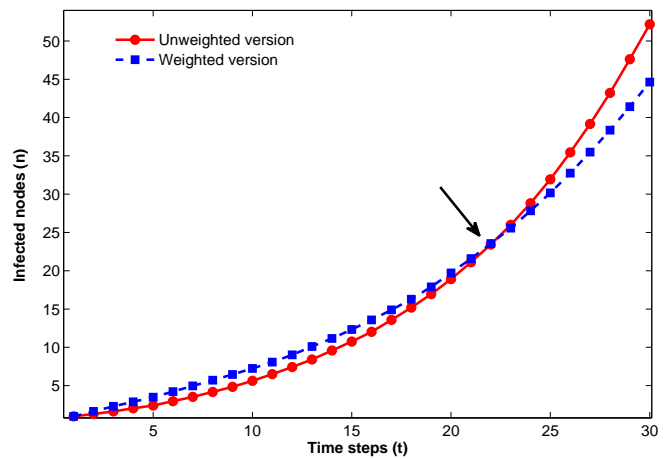}
\caption{(Color online) The number of infected nodes as a function
of time $t$ in the short message communication network. Here we use
the Susceptible-Infected (SI) model to simulate information
spreading dynamics. The result is the average from selecting the node
whose out-strength is larger than $100$ as the start node. In the
weighted version, it is assumed that the probability for a node $i$
to pass on the information to its neighbor $j$ in one time step is
given by $P_{ij}=xw_{ij}$, and $x=0.003$ \cite{Onnela_2007}. We also
compare the above result with that obtained in an unweighted version
(all the tie weights are considered equal) of the short message
communication network using the same average spreading rate.}
\label{spreading}
\end{figure}

We now study a special spreading process in which a rich strength
node has the initializing information, since a node with a higher
strength has a higher probability to transfer information to others.
On the large scale, information is transferred faster in the
unweighted network than the weighted version, which is essentially
coincident with the result in \cite{Onnela_2007}, because the rich
isolated multipolarization phenomenon means that there are many
bottlenecks for global information spreading in the weighted network
[Fig. \ref{multipolarization}]. On the contrary, the weighted
version has a faster spreading speed than the unweighted version on
the small scale [Fig. \ref{spreading}]. This finding implies that
for real information spreading, previous works may underestimate the
spreading speed of local information. However, the pairwise
interaction leads to the result that most messages are only sent to
the most closest person, so it is difficult for a pair of intimate
individuals to transmit their common local information to a large
range (to be a global information). In summary, the pairwise pattern
promotes local information spreading among a pair of friends and in
contrast also suppresses global information spreading.

\subsection{Impact on human dynamics}
We also confirm a strong correlation between the pairwise
conversation and human dynamics. There are two important time
periods with which to describe human dynamics: the waiting time and
the interevent time. The waiting time $\tau_{i}$ is the interval
between receiving and then sending a short message for user $i$, and
the interevent time $t_{i}$ is the time interval between sending two
consecutive messages. The effect of bursts on hindering information
propagation has been well studied \cite{Barabasi_2005, Vazquez_2007,
Iribarren_2009, Min_2011}, yet there are few works to study the
effect of the conversation mechanism within human crowds
\cite{Wu_2010}. For the mobile phone call data, recently Miritello
\emph{et al.} found that group conversations enhanced
 local information spreading, because the interaction among
individuals can trigger more information spreading behaviors
\cite{Miritello_2011}.

\begin{figure}[htbp]
\centering
\includegraphics[width=0.5\textwidth]{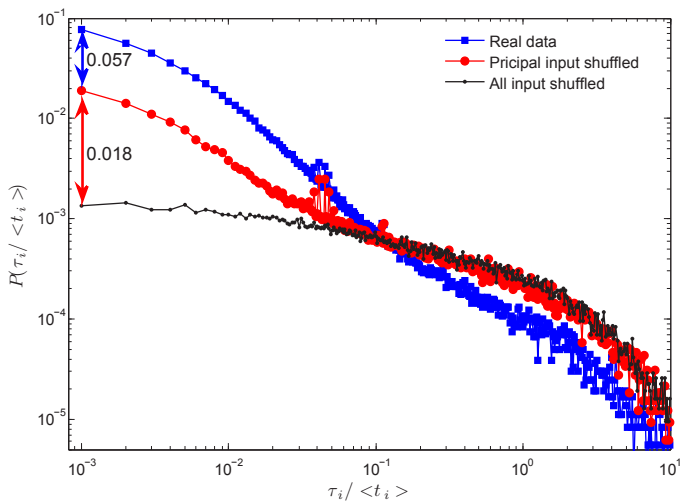}
\caption{(Color online) The distribution of the waiting time
rescaled by the average of the interevent time. The waiting time
$\tau_{i}$ is the interval between user $i$ receiving a message and
then sending a short message to another user. The interevent time
$t_{i}$ is the time interval between sending two consecutive
messages for user $i$. Here $\left<t_{i}\right>$ represents the
average of the interevent time $t_{i}$ of the active user $i$. The
active users are the persons who send more than five short messages.
For the case of ``principal input shuffled'', we only shuffle the
receiving time of short messages from the one major active
communication partner. For the case of ``all input shuffled'', we
shuffle the receiving time of short messages from all the users. }
\label{tauij}
\end{figure}

In this study, we verify again that $\tau_{i}$ depends on the group
conversations in Fig. \ref{tauij}. Using the similar framework in
\cite{Miritello_2011}, we compare the results of $\tau_{i}$ for only
shuffling the receiving time of short messages from the one major
active communication partner and shuffling all the receiving time.
Both most short messages being sent to one particular person [Fig.
\ref{rsmax}] and the weighted reciprocal coefficients being high
[Fig. \ref{reciprocity}] indicate that there are no frequent
cascading information spreading in the short message dataset.
Therefore, the event that user $i$ receives the information from
user $m$ ($m\to i$) is not a trigger for $i$ to communicate with
other people like user $n$ ($i \to n$ and $m\neq n$). However, we
find that the pairwise interaction is the basic pattern of group
conversations ($i \to j$, and then $j \to i$). The pairwise
interaction is the most significant impact on human dynamics, for it
can greatly reduce the waiting time between a pair of close friends
[Fig. \ref{tauij}].

\section{Conclusion and Discussion}
In summary, we have found a strong pairwise interaction pattern in
the weighted communication network by analyzing the network topology
and interaction strengths. The pairwise interaction, which has not
been reported in other communication datasets by unweighted network
analysis, is the basic conversation pattern among individuals and it
has a significant impact on human dynamics of communication
behaviors: reducing the waiting time in electronic person-to-person
communications.

Our finding suggests that the short message service promotes local
information spreading and slows down global information cascading
spreading. Basically, anyone can transfer information to anyone, for
human society has a famous small-world property \cite{Small_wolrd1,
Small_wolrd2}. However, interaction strengths play distinct roles
for information spreading \cite{Onnela_2007}. It is believed that
weak ties have a strong impact for long-range information spreading
and strong ties are significant for local information spreading
\cite{Granovetter_1973}. In the short message dataset, most
information is local (especially tends to happen between a pair of
users), and the global cascading spreading is not frequent. Our
results are coincident with the fact that the short message
communication system is not designed for spreading global
information such as rumor and news, but is self-organizing to
support us to spread or exchange local information in daily life.

The collective communication behaviors of all the customers in a
company belonging to the same mobile phone operator have been
analyzed based on the weighted network theory. Our work is helpful
for mobile phone operators to design new service plans and tariff
strategies. For example, it is valuable for a mobile phone operator
to provide a special tariff for a pair of users with frequent
communication relationships. Furthermore, because the number of the
user's friends obeys a power-law distribution \cite{Barabasi_1999},
it is a remarkable fact that most users have only one major active
communication partner ($70\%$ messages are sent to the same person).
Therefore, our findings also can be used to optimize communication
services.

For a self-organizing complex system (e.g., the short message
communication system in this study), it is well known that ``the
whole is greater than the sum of its parts''
\cite{More_is_different}, so dividing a large-scale complex system
into multiple sub-systems (or individuals) is not helpful for the
comprehensive understanding on the whole complex system
\cite{Newman_survey}. The pairwise interaction pattern suggests that
human dynamics not only depend on individuals' rhythm
\cite{Jo_circadian}, but also are strongly affected by the
complicated interplay among intimate friends. Nevertheless, our work
is a step in an ongoing effort to bridge the gap between individual
microscopic interactions and macroscopic social systems.

\begin{acknowledgments}
This work was supported by the PolyU Postdoctoral Fellowships Scheme
(G-YX4A) and the Hong Kong University Grants Council General
Research Fund (PolyU 5300/09E). X.-K.X. also acknowledges the
support of the National Natural Science Foundation of China
(61004104, 61104143).
\end{acknowledgments}

\end{document}